\documentclass[12pt]{article}
\usepackage[pdfstartview=FitH,colorlinks=true,linkcolor=blue,anchorcolor=red,citecolor=magenta,urlcolor=blue]{hyperref}
\usepackage{amsmath,amssymb,titling,authblk}
\usepackage{amscd}
\usepackage{tcolorbox}
\usepackage{amsfonts}
\usepackage{amstext}
\usepackage{amsthm}
\usepackage{bbm}
\usepackage[normalem]{ulem}
\usepackage{appendix}
\usepackage{bbold}
\usepackage{yfonts}
\usepackage{epsfig}
 \usepackage{microtype}
 \usepackage[utf8]{inputenc}
\usepackage[T1]{fontenc}
\usepackage {color}
\usepackage{bbold}
\usepackage{latexsym}
\usepackage{soul}
\setstcolor{blue}
\usepackage{cite}
\allowdisplaybreaks[4]
\numberwithin{equation}{section}

\definecolor{verde}{cmyk}{.83,.21,1,.08}

\newtheorem{theorem}{Theorem}

\newtheorem*{proof*}{Proof}

\advance\hoffset by -1.1truecm
\advance\voffset by -1.0truecm
\newcommand{\be}{\begin{equation}}
\newcommand{\ee}{\end{equation}}
\newcommand{\beqa}{\begin{eqnarray}}
\newcommand{\eeqa}{\end{eqnarray}}
\newcommand{\eqn}[1]{(\ref{#1})}

\newcommand{\del}{\partial}
\newcommand{\dd}{\mathrm d}
\newcommand{\ii}{\mathrm i}
\newcommand{\e}{\mathrm e}
\newcommand{\R}{\mathbbm R}
\newcommand{\Z}{\mathbbm Z}

\newcommand{\x}{\mathbf x}

\numberwithin{equation}{section}

\newcounter{appendice}

\usepackage{lmodern}

\begin{document}

%%%%%%%%%%%%%%%%%%%%

%\title{\boldmath The $\mathfrak{so}(2,2)$ Yang-Mills extension of JT-Gravity via the Virasoro-Kac-Moody semidirect product}

\title{{\boldmath A $\mathfrak{so}(2,2)$  extension of JT gravity via the Virasoro-Kac-Moody semidirect product}}

\author[a,b]{Goffredo Chirco \thanks{goffredo.chirco@unina.it}}
\author[a,b]{Lucio Vacchiano \thanks{lucio.vacchiano@unina.it}}
\author[a,b]{Patrizia Vitale\thanks{patrizia.vitale@na.infn.it}}

\affil[a]{Dipartimento di Fisica `Ettore Pancini', Universit\`a  di Napoli Federico II, Via Cintia 80126,  Napoli, Italy}
\affil[b]{INFN, Sezione di Napoli, Italy}

\date{}

\maketitle

\vspace{-2cm}
\begin{abstract}We consider a bulk plus boundary extension of Jackiw–Teitelboim Gravity (JT) coupled with non-abelian gauge fields. The generalization is performed in the Poisson Sigma Model formulation and it is derived as a dimensional reduction of the AdS$_3$ Chern-Simons theory with WZW boundary terms. We discuss the role of boundary conditions in relation to the symmetries of the boundary dynamics and we show  that the boundary action can be written in terms of coadjoint orbits of an appropriate Virasoro-Kac-Moody group. We obtain a Schwarzian action and interaction terms with  additional edge modes that match the effective low energy action of recent SYK-like tensor models. 

\bigskip
\textbf{keywords:} {\it 2d dilaton gravity, asymptotically AdS, JT model, 
Poisson Sigma-Models, Virasoro-Kac-Moody coadjoint orbits, Schwarzian action.} 
\end{abstract}
\bigskip
\section{Introduction}
Two-dimensional dilaton gravity has proved to be a powerful tool in describing black hole physics at the classical and quantum level as an effective description of the near-horizon dynamics obtained from dimensional reduction~\cite{GRUMILLER2002327, BANKS1991649, IKEDA1994435}. Among these models, Jackiw-Teitelboim gravity (JT)~\cite{Jackiw:1982hg,TeitelboimGHS2D,Jackiw:GTGON} has attracted much attention recently, because of its holographic duality with the one-dimensional Sachdev–Ye–Kitaev model (SYK), which provides a concrete and solvable realization of the (near) $\operatorname{AdS}_2 /\operatorname{CFT}_1$ correspondence \cite{PolchinskiADS, maldacenaConformal, MertensBackreaction, CvetiDictionary}. In the JT/SYK case~\cite{SachdevRN}, the duality lies in a shared one-dimensional Schwarzian dynamical sector, which describes the low-energy/strong-coupling regime of the SYK model \cite{SachHolo}, as well as the boundary dynamics induced by the Gibbons-Hawking-York (GHY) term in the JT theory. 
%Understanding the Schwarzian action as a particular coadjoint orbit of the Virasoro group has been of great value in developing a deeper comprehension of the solvability of this model~\cite{bb}. 
In the case of JT gravity, the reduction of the GHY term to the Schwarzian action at the boundary has been computed explicitly (see~\cite{Mertens} and references therein) to give
\begin{equation}
\label{schwder}
    S_{JT}|_{\partial \Sigma} \simeq \int_{\partial \Sigma} \Phi_b\{F(\tau), \tau \}_{\mathcal{S}} \, d\tau
\end{equation}
where $\tau$ is a coordinate  parametrizing  the boundary $\partial\Sigma\simeq S^{1}$, 
$F$ a diffeomorphism of the boundary, $\Phi_b$ the boundary value of the dilaton when adopting Dirichlet boundary conditions, and
\begin{equation}\label{Sder}
   \{F(\tau), \tau \}_\mathcal{S} = \frac{F'''}{F'} - \frac{3}{2}\left( \frac{F''}{F'}\right)^2 
\end{equation} 
the Schwarzian derivative~\cite{schwarz1972gesammelte, OvisenkoSchwarzian}.
The latter transforms as a CFT stress-energy tensor under reparametrization of the circle, and it is invariant under Moebius transformations. %The presence of a boundary explicitly break the diffeomorphisms symmetry. 

On the boundary of the near $AdS_2$ (near-horizon) geometry, the Schwarzian derivative encodes the notion of extrinsic curvature~\cite{OvisenkoSchwarzian, Iliesiu} while being associated with the breaking of conformal symmetry  down to a global SL$(2,\mathbb{R})$ both in JT and SYK-like models~\cite{maldacenaConformal}. In the first case, the conformal symmetry is 
spontaneously broken by the presence of a boundary and explicitly broken in the passage from pure AdS to near AdS via
the coupling to the dilaton, while, in the second case, the breaking reflects a finite-temperature effect. 
Besides the duality, the beauty and power of such a framework lies in the generality of the symmetry-breaking scheme in low dimensional gravitational theories with a boundary (see also \cite{balachandran1995edgestatescanonicalgravity,Carlip95, BTZ92}).
%The Schwarzian action can in fact be derived as the geometric action associated with a specific coadjoint orbit of the Virasoro group. The method of coadjoint orbits has revealed to be a fruitful way to model the boundary action of JT like gravitational theories for the following reason: 

The presence of a boundary breaks the asymptotic group of symmetry $\mathcal{G}$   into some reduced symmetry group $\mathcal{H}$ which depends on the choice of boundary conditions. The boundary dynamics is governed by fields belonging to the quotient space $\mathcal{G}/\mathcal{H}$ which can be identified with an appropriate coadjoint orbit of $\mathcal{G}$.
%, with $\mathcal{H}$ playing the role of residual symmetry. 
In the case of JT gravity, one has $\mathcal{G} = \operatorname{Diff}(S^1)$  and $\mathcal{H} = \operatorname{SL}(2,\mathbb{R})$. The boundary %JT 
action can then be written as the geometric action on $S^1$ associated with the coadjoint action of a diffeomorphism $F$ over an element of the Virasoro dual algebra and, with a suitable choice of this element, the action in \eqref{schwder} can be exactly derived in purely geometric terms. We refer to Section \ref{SchActionCoad} and Appendix \ref{appendix} for a detailed review of the derivation.

The generality of the coadjoint orbits method provides a powerful tool for investigating extensions of the boundary JT/SYK dynamics starting from generalizations of the symmetry group of the dual models.   
Indeed, ever since the JT/SYK correspondence has been pointed out, much work has aimed at expanding the range of potential holographic relationships in 2d gravity~\cite{Grumiller,Menagerie, Mertens:2020hbs, Ecker:2021guy, Ecker:2023sua}.  
%Since the SYK/$\operatorname{AdS_2}$ correspondence has been pointed out, a great deal of interest has arisen in understanding possible extensions of such a duality. 
Numerous variants and generalizations of the SYK model have been proposed in recent years \cite{Gross1,Gross2,WittenSYK,Yoon,DavisonSYK, NarayanYoon, MaldacenaFuSYK},  raising the natural question of whether an extended duality with  generalized  JT models would exist. In this sense, modifications of JT gravity have been introduced following diverse paths, such as relaxing the boundary conditions \cite{GrumillerMenagerie}, considering higher spin extensions of the theory \cite{AlkalaevHS,GrumillerHS,Grumill}, or coupling the gravitational sector with additional degrees of freedom, such as non-abelian gauge fields \cite{Grumill}. %Therefore a natural question to rise is whether or not these two classes of generalized model are dual in the same sense the SYK model and the JT-Gravity model are. 
Among the various generalizations of the SYK model, on the other side of the correspondence, tensor extensions in particular have been shown to exhibit a broken $\mathfrak{diff}(S^1)\ltimes \Hat{\mathfrak{g}}$ symmetry in the strong coupling limit, with $\mathfrak{g}$ the Lie algebra of some internal symmetry by which the tensor extension is realized \cite{Yoon} and the hat indicating, as in the rest of the paper, the associated loop algebra.
\textcolor{black}{One main motivation for tensor generalizations of SYK models  lies in the fact that, while preserving solvability in the large N limit, their effective dynamics turns out to be a generalization of Schwarzian dynamics, without  requiring disordered averaging. As argued in \cite{WittenSYK}, the possibility of establishing holographic relations without necessarily averaging over disorder is a remarkable feature of these models because it allows for a more direct correspondence between the tensor models degrees of freedom and those of gravitational theories in low dimensions.  As a further motivation for the introduction of tensor
models we also mention  the idea of generalizing to higher dimensions the  correspondence between matrix models and two-dimensional geometries \cite{Gurau:2009tw, Bonzom_2011}.}

The emergent symmetry of tensor extensions of  SYK models is  generally characterized by the semidirect product
of the group of reparametrisations of $S^1$ with an affine Kac-Moody algebra. Therefore, a natural question is whether a bulk gravitational theory exists whose boundary action is written in terms of coadjoint orbits of such semidirect product.  

In this work, we propose a  bulk plus boundary theory which reproduces a broken $\mathfrak{diff}(S^1)\ltimes \Hat{\mathfrak{g}}$ symmetry at the boundary and can be regarded as a gauge extension of the JT gravity model, in the sense indicated in \cite{Grumill}. The way the extension is realized differs from what has been proposed so far in the literature, in many respects. In \cite{Grumill}, for instance, the authors consider an $\mathfrak{sl(2,\mathbb{R})}$-BF theory for the gravitational sector together with an additional $\mathfrak{g}$-BF theory for the Yang-Mills part, so that the symmetry of the model is given by the direct sum $\mathfrak{sl(2,\mathbb{R})} \oplus \mathfrak{g}$. However, the proposed bulk duals, to date, do not have a clear gravitational interpretation and the role of the additional degrees of freedom extending the JY/SYK correspondence remains unclear. 

\textcolor{black}{We propose a class of extensions of JT %with a clear purely gravitational interpretation
where the presence of additional degrees of freedom is fully ascribed to the request that the 2d theory be obtained from a dimensional reduction of a Chers-Simons theory describing a maximally symmetric space in three dimensions, which includes a BTZ black hole solution (see e.g.\cite{PhysRevD.48.3600, Verlinde, Origin}). This choice drastically reduces the space of possible duality relations, but at the same time allows the additional degrees of freedom to be clearly interpreted as Kaluza-Klein modes emerging from the above dimensional reduction. Notice that, although  starting from three-dimensional theories limits the options on the possible symmetry groups for the corresponding 2d theories,  the choice of the group does not unambiguously determine a gravitational dual for a given extension of SYK, as much depends on the boundary conditions. Consequently, when a symmetry group is fixed with a given criterion, i.e. compatibility with a pure gravity theory in 3d, as many duality relations can be realized as the allowed choices of boundary conditions. 
}
 
% \textcolor{blue}{ In the present work we embed the gravitational $\mathfrak{sl}(2,\mathbb{R})$ sector of JT gravity within a $\mathfrak{so}(2,2)$ Poisson Sigma Model. Such a model naturally relates to the one-dimensional effective dynamics of the SYK tensor model. Both models describe the physical modes associated with the breaking of Diff$(S^1)\ltimes \mathcal{LG}$, where $\mathcal{LG}$ is some Loop Group associated with the partial breaking of internal symmetries at the boundary. The main advantage of such a construction is that, depending on the boundary conditions, one can in principle derive a family of boundary theories and therefore establish an enlarged correspondence relation with a family of SYK-like tensor models with different global symmetries. Moreover, the bulk theory that we analyse has a direct relation with 3d theories of pure gravity like the AdS$_3$ Chern-Simons theory. }(\textcolor{magenta}{Questo paragro risulta come una ripetizione dopo le tue aggiunte sopra e sotto. Possiamo toglierlo e integre alcuni dettagli nel testo che segue. (?)})

\textcolor{black}{Our construction can be summarized  by the following steps. Building on the previous work which relates JT gravity, BF-theory and linear Poisson Sigma models (see e.g. \cite{Iliesiu} and references therein), we first derive the 
 Poisson Sigma model with linear Poisson bracket of $\mathfrak{so}(2,2)$ type from the dimensional reduction of the $\mathfrak{so}(2,2)$ Chern-Simons theory with WZW boundary terms. This  is a purely gravitational theory with a boundary dynamics, as well as JT gravity is in 2d. The two dimensional theory obtained  in this way is an extension of JT gravity which also includes additional non-abelian gauge fields both in the  bulk and at the boundary.
%  To this end, since the $\mathfrak{so}(2,2)$ algebra is isomorphic to $\mathfrak{sl}_{L}(2,\mathbb{R})\oplus\mathfrak{sl}_{R}(2,\mathbb{R})$, we take the chiral basis 
% \begin{equation}
% \label{so22}
%     [L_i, L_j] = c^k_{ij} L_k, \qquad [R_i, R_j] = c^k_{ij} R_k, \qquad [L_i,R_j] = 0
% \end{equation}
% with $c^k_{ij}$  structure constants of $\mathfrak{sl}(2, \mathbb{R})$, and we rotate it into  
% \begin{equation}
% \label{so22lj}
%     [J_i, J_j] = c^k_{ij} J_k, \qquad [L_i, L_j] = c^k_{ij} L_k, \qquad [L_i,J_j] = c^k_{ij}J_k,
% \end{equation}
% with $J_i = R_i + L_i$.
% Notice that the $L_i$-generators close an invariant sub-algebra under the action of the $J_i$ generators. The goal of of this redefinition is to interpret the $\mathfrak{sl}(2,\mathbb{R})_{J}$ subalgebra as that associated to the JT gravity sector and the $\mathfrak{sl}(2,\mathbb{R})_{L}$ as the Lie algebra of an additional Yang-Mills sector.
The choice of $\mathfrak{so}(2,2)$ is  the simplest non trivial choice which makes it possible  to understand the model as a dimensional reduction of a purely gravitational 3d theory of a kinematical space-time, in the sense of \cite{Kine}, given that SO$(2,2)$ is the isometry group for AdS$_3$}. 

\textcolor{black}{The WZW boundary term for the 3d theory reduces to   a boundary term in the form of a Casimir function for the $\mathfrak{so}(2,2)$ algebra in the 2d theory. Together with an appropriate choice of boundary conditions, this leads to a boundary dynamics governed by the $\mathfrak{diff}(S^1)\ltimes \hat{\mathfrak{g}}$ symmetry breaking. The boundary Casimir action is then identified with the action associated with coadjoint orbits of the Virasoro-Kac-Moody semidirect product. We explicitly compute the Kac-Moody terms and investigate their contribution as corrections to the nearly extremal entropy in JT. In particular, we discuss how the obtained results do not depend on the specific parameterization choices for edge fields, such as the highest weight gauge widely used in the literature~\cite{Grumill}.
}

\textcolor{black}{The work is organised as follows. 
In Section \ref{sec2}, we recall the formulation of JT gravity as a 2d SL(2,$\R$)-BF theory and its derivation from a linear SL(2,$\R$)- Poisson Sigma Model (PSM). As we shall see, the latter formulation allows for a more natural introduction of boundary terms, differing  from the BF action for a boundary contribution. Moreover, as we shall comment in the Conclusions, lends itself to generalizations. In the PSM framework, we review the construction of dynamical boundary actions in terms of a Casimir function and their relation with coadjoint orbits of infinite-dimensional groups. Finally we describe the  dimensional reduction from 3d CS to our 2d model. The details on  the Kirillov method of coadjoint orbits of infinite dimensional Lie groups are given in Appendix \ref{appendix}. Sections  \ref{Sec3} and \ref{bhe} contain the original results of the paper. In Sec.  \ref{Sec3} we discuss the extension of the JT model via a $\mathfrak{so}(2,2)$-PSM and the need for the Virasoro-Kac-Moody semidirect product. Here, we also provide the explicit computation of the coadjoint orbits, including the Kac-Moody contributions. In section~\ref{bhe} we compute the leading order entropy in the Euclidean theory and compare it with the SYK results. We close in Section~\ref{Sec4} with a summary of our results. We also further motivate the choice of working with a PSM with the indication of some future directions of research. }

\section{2d JT Gravity as PSM/BF Theory}
\label{sec2}
Before discussing the details of the $\mathfrak{so}(2,2)$-Poisson Sigma Model (PSM), it is worth to briefly review 
the topological gauge theory description of JT gravity in the first order formulation, the way it reduces to a BF model~\cite{Fukuyama, Isler} or equivalently to linear PSM~\cite{Schaller} (see also \cite{Mertens} for a review). Upon reviewing the formalism, we will consider a two-dimensional manifold ${\Sigma}$ with a boundary $\partial \Sigma$ and focus on the form of the boundary terms in the description of BF/PSM.

\subsection{JT gravity in the BF/PSM formalism}
%It is well known that  JT gravity can be written as a BF Theory and as a particular case of PSM.  

The Jackiw Teitelboim two dimensional gravity theory is defined by the action   \cite{Jack} 
\begin{equation}\label{sjt}
    S_{JT}[g, \Phi]  = -\frac{1}{16\pi G_N}\int_{\Sigma} \dd^2 x \sqrt{-g}\,\Phi(R+2)+ \int_{\partial \Sigma} \dd u{\sqrt{h}(K-1)\Phi_b }
\end{equation}
where $\Phi$ is the dilaton and $R$ the scalar curvature of the bulk, while  $\Phi_b$,  $K$ and $\dd u\sqrt{h}$ respectively refer to the restriction of the dilaton field on the boundary, the extrinsic curvature of $\partial \Sigma$ and the boundary volume form. The second term in \eqref{sjt}, the Gibbons-Hawking-York (GHY) boundary term~\cite{GHY}, is included to make the metric variational problem well defined. The bulk action in \eqn{sjt} can be explicitly derived from the near-horizon and near-extremality limit of the Reissner-Nordstrom solution in four dimensions, via dimensional reduction %from four to two dimensions 
\cite{SachdevRN,NayakRN}. %The boundary term is the ordinary Gibbons-Hawking-York  term \cite{GHY}. 
The model is topological in the bulk and the dilaton field just acts as a Lagrange multiplier fixing the value of the curvature. Although there are no propagating degrees of freedom in the bulk, under suitable choices of boundary conditions, the boundary action is dynamical. 

Let us first focus on the bulk. The bulk JT action can be formulated, in a first-order formalism, as a 2d topological SL$(2,\R)$ BF theory  
%As a first step, we construct an $\mathfrak{sl}(2,\mathbb{R})$-BF theory.
%The $\mathfrak{sl}(2,\mathbb{R})$ algebra is given by 
%\begin{equation}
%    [J_a,J_b] = \epsilon^c_{\,ab}J_c,
%\end{equation}
%where $\epsilon^c_{ab}$ is the contraction of the metric tensor $\eta_{ab}=\text{diag}(1,1,-1)$  and the ordinary Levi-Civita tensor. 
%What we intend to do is ultimately identifying two of the three $e^k_\mu$ fields with zweibeins and one component of $\Phi$ with the JT dilaton field. 
%Given a two dimensional manifold $\Sigma$, which we assume is a disk, we can write 
\begin{equation}\label{SBF}
    S_{BF}[B,A] =  \int_{\Sigma} \operatorname{Tr}( B F)
\end{equation}
where  $F$ is the curvature of the  Lie algebra valued connection 1-form $A = A^{a}_{\mu}\,dx^{\mu} J_{a}$, $B=B^a\,{J_a}$ a Lie algebra valued  scalar field and $J_a$ the ${\mathfrak{sl}(2,\R)}$ Lie algebra generators.  
Varying the action with respect to $B$ leads  to the equation of motion $F=0$, which is equivalent to saying that the connection is locally trivial (pure gauge), that is
\begin{equation}
\label{eomf}
    A = g^{-1}dg
    \end{equation}
where $g$ is an element of the gauge group.

The  bulk term in the action \eqn{sjt} is recovered as follows. One first identifies the components of the gauge connection with the zweibein forms on $\Sigma$ and the Lorentz (or spin) connection (Cartan variables), that is $A^{0,1,2}_\mu=( e^{0}_{\mu},e^{1}_{\mu}, \omega_{\mu})$~\cite{Fukuyama}. The dilaton field is encoded in one component of the field $\Phi=B^{{2}}$, while the other two, $B^{0,1}$ define Lagrange multipliers imposing the condition of zero torsion for the spin connection
\begin{equation}
\label{vielpost}
    de^a + \omega\, \epsilon^{ab} \wedge e_b = 0.
\end{equation}
Equation \eqref{eomf}, together with the assumption that the zweibein  is invertible, yields 
\begin{equation}
\label{spincon}
    {\omega}_{\mu}= -e^{-1}\, \epsilon^ {\gamma \delta} \,\partial_{\gamma} e^{k}_{\delta}\, e_{k\mu} ,
\end{equation}
where $g_{\alpha \beta} = e^{h}_{\alpha}\,e^{k}_{\beta}\, \eta_{hk}$, with $ h,k =0,1$, and $e = \operatorname{det}\{ e^{k}_{\mu} \}$. Finally, the spin connection being defined  as ${\omega_{\mu}}^{ab} =\omega_{\mu}\epsilon^{ab}$, we can identify the curvature as ${F_{\mu \nu}}^{2} = R_{\mu \nu}$, and the torsion ${F_{\mu \nu}}^k = {T_{\mu \nu}}^k$, for $k=0,1$. The equations of motion $R_{\mu \nu}$ = 0 together with Eq. (\ref{spincon})  lead to 
\begin{equation}
    \sqrt{-g} \,\Phi\left(\, R(g_{\alpha \beta}) + 2\,\right) = 0 .
\end{equation}
Therefore, the on-shell BF action reduces to the bulk term of the JT action $S_{JT}[g,\Phi]$. The on-shell connection is pure gauge and the dynamics of the dilaton corresponds exactly to a gauge transformation that preserves the form of $A$, that is $B$ on-shell is the stabilizer of $A$. 

%The BF formulation of JT gravity neglects the boundary term. In order to have a well-defined variational principle and, ultimately, to reproduce the GHY term of JT gravity, it is necessary to introduce an additional boundary contribution to the BF action. 
Equivalently, one can write the 2d BF theory as a linear Poisson sigma model (PSM) in terms of real fields $(X, A)$, with $X : \Sigma \to M$ the usual embedding map and $A \in \Omega^1(\Sigma, X^*
(T^*M))$ a one-form on $\Sigma$ taking values in the pull-back of the cotangent bundle
over $M$. The action of the general PSM is given by ($  i, j=1, \dots,\mbox{dim} M$)
\begin{equation}
S(X,A)= \int_{\Sigma}  A_i{\wedge} dX^i  + \frac{1}{2} \Pi^{ij}(X) A_i\wedge A_j, 
\end{equation}
where  $dX \in \Omega^1 (\Sigma, X^*(TM))$  and the contraction of covariant and contravariant indices
is  induced by the natural pairing between $T^*M$ and 
$TM$,   
yielding  a two-form on $\Sigma$. To make contact with the BF formulation of JT gravity described above, one has to  consider  a linear Poisson tensor of  Lie algebra type
\begin{equation} \label{PT}
    \Pi^{ij}(X) = f^{ij}_{k}X^{k}
\end{equation}
with   $f^{ij}_{k}$ the structure constants of the $\mathfrak{sl} (2,\mathbb{R})$ Lie algebra. 
In particular, integrating by parts the  linear PSM action with Poisson tensor given by \eqn{PT} we obtain the BF action \eqn{SBF} plus a boundary term: 
\begin{equation}
\label{psb}
    S_{PSM} = S_{BF} - \int_{\partial \Sigma} X^{i}A_{i}.
\end{equation}
The variation of the action with respect to $X$ and $A$ yields
\begin{equation}\label{var}
    \delta S_{PSM} = \int_{\Sigma}(E.L.)\, \delta X^{i} + (E.L.)\,\delta A_{i} - \int_{\partial \Sigma} \delta X^{i}A_{i}.
\end{equation}
with Euler Lagrange equations (E.L)
\begin{equation}
    \label{EOMXA}
    {D}_A A =0,\,  \qquad dX + [X,A]=:\delta_X A =0,
\end{equation}
where ${D}_A$ denotes the covariant derivative with respect to the gauge connection $A$. The first equation implies that A is pure gauge 
\begin{equation}
\label{puregauge}
    A = g^{-1} dg, 
\end{equation}
with $g: \Sigma \to \text{SL}(2,\R)$, while the second equation states that the on-shell $X$ field is a stabilizer of $A$. As suggested in \eqref{EOMXA}, the dynamics of the dilaton corresponds to an infinitesimal gauge  transformation that preserves the form of $A$ along $X$ on-shell. In particular, one can check that the boundary term in \eqref{psb} is such that the gauge invariance restricts to the gauge transformations that satisfy $\delta_g A|_{\partial \Sigma} =0$, which is exactly the equation of motion for the $X$ field when restricted at the boundary. The reparametrization symmetry Diff$(S^1)$ is broken and this breaking  is responsible for the rise of dynamical boundary degrees of freedom (see \cite{Origin} and references therein). 

Now the presence of the boundary terms in \eqref{var} requires fixing the boundary values of the fields to have a well-defined variational principle. However, this choice would prevent any boundary dynamics. Alternatively, adding an extra boundary term allows for more general boundary conditions and preserves the variational principle. %In order to reproduce the boundary dynamics induced by the GHY term of the JT action it is necessary to introduce an additional boundary term. 
\textcolor{black}{The PSM framework suggests a natural choice in this sense, consisting in the Casimir function $X^2$}.\footnote{Any smooth function of $X^2$ is a Casimir function for the Poisson algebra $\mathcal{F}(\Sigma)$ with  Poisson bracket \eqn{PT}, it being $\{X^2, -\}=0$ }
%, since we need to trace out the algebra
Indeed one can write 
\begin{equation}
 \label{boundcas}
     S_{\,(\Sigma + \partial \Sigma)} = S_{PSM} + \frac{1}{2} \int_{\partial \Sigma} X^{i}X_{i}\, \dd u\, \, , 
 \end{equation}
with $\dd u$ the integration measure  over $\partial\Sigma$, with the condition that
\begin{equation}
\label{boundcond}
    X_{i}|_{\partial \Sigma}\, \dd u = A_{i}|_{\partial \Sigma}.
\end{equation}The presence of the extra boundary term in \eqref{boundcas} allows for more general boundary conditions and   preserves the  variational principle. The condition in \eqref{boundcond}, together with the fact that $A$ is pure gauge on-shell, and the required continuity of the fields at the boundary, leads  to a boundary action which describes the dynamics of a particle on a group manifold, in this case on SL$(2,\R)$. Indeed, by continuing Eq. \eqn{puregauge} to the boundary, the on-shell boundary action reads\footnote{One might wonder  why not to include the Casimir function directly  in the bulk action. The reason, as explained in \cite{Strobl}, is that the addition of a Casimir term in the bulk would break the topological nature of the model, resulting in a theory which is equivalent to a Yang-Mills in the bulk, which is not what we want.
}
\begin{equation}
\label{POGA}
    S|_{\partial \Sigma} = \int_{\partial \Sigma} \operatorname{Tr}(g^{-1}g')^2 \frac{1}{u'} \dd \tau,
    \end{equation}
that is a particle on a group action where now $g$ is an element of the loop group $\mathcal{L}\text{SL}(2,
\R)$.   

At this stage, it is not obvious that the boundary quadratic term derived above encodes the same boundary dynamics induced by the GHY term in JT gravity. However, it is possible to show that the boundary dynamics of the particle on the group manifold is related with the Schwarzian action.
In the following section we review the elements of such a correspondence. Thereby, in Section~\ref{Sec3}, we propose a generalization of this result for the case of a $\mathfrak{so}(2,2)$ PSM.

\subsubsection{Schwarzian Action from Coadjoint Orbits}
\label{SchActionCoad}
As we already discussed in the previous section, the presence of a boundary action \eqref{POGA} breaks the reparametrisation invariance associated with Diff$(S^1)$ to global SL(2,$\R$) on the boundary, while the gauge symmetry is automatically mod-out since the equations of motion \eqref{EOMXA} require the $A$ connection 1-form to be pure gauge. Therefore, the particle on a group action shows a global $\operatorname{SL}(2,\mathbb{R})$ symmetry, modulo gauge transformation which are trivial on the boundary~\cite{Valach}. Accordingly, the dynamical boundary action is reduced to the coset space Diff$(S^1)/\operatorname{SL}(2,\mathbb{R})$ defined by the aforementioned symmetry breaking pattern. 

Such a reduction can be performed by means of Kirillov's coadjoint orbit method (see Appendix \ref{appendix} for a detailed review). In brief, coadjoint orbits of the Virasoro group over the dual Virasoro algebra naturally realize  homogeneous spaces
$\operatorname{Diff(S^1)}/\operatorname{Stab}(b)$, where $\operatorname{Stab}(b)$ indicates the little group associated with the dual element $b$ over which the coadjoint action is computed. In the present case, we want $\operatorname{Stab}(b)$ to be the global residual SL$(2,\mathbb{R})$. In particular, being $b=(f, t)$ an element in the dual of the Virasoro algebra and $\phi$ a finite diffeomorphism in the Virasoro group, we can compute the coadjoint action (see eq. \eqref{caodVir}) as follows 
\begin{equation}
\label{coadaction}
\tilde{b}=Ad^*_{\phi^{-1}} (f(\tau),t) = \left({\phi'}^2 f(\tau) - \frac{t}{12}\{\phi(\tau),\tau \}_{\mathcal{S}}, t\right)\, \in \mathfrak{diff}^*(S^1) 
\end{equation}
For $f(\tau) =- \frac{tn^2}{24}$ one gets that $
\operatorname{Stab}(b) = \operatorname{SL}^{(n)}(2, \mathbb{R})$ and the homogeneous space defined by the coadjoint orbit is $\operatorname{Diff(S^1)}/\operatorname{SL}^{(n)}(2,\R)$,
where $\operatorname{SL}^{(n)}(2, \mathbb{R})$ denotes the $n$-fold cover of $\operatorname{SL}(2, \mathbb{R})$ (see e.g. \cite{Valach}).

Now, a natural action functional on the coset space is given by the pairing of $\tilde{b}$ with an element $\xi \in \mathfrak{diff}(S^1)$, which gives  \cite{Barnich_2018}
\be
\label{barnich}
\langle \tilde{b}, \xi \rangle= \int_{S^1} \tilde{b}(\tau) \xi(\tau) \mathrm{d} \tau\, .
\ee
The latter can be identified with the reduction of the on-shell boundary action 
\begin{equation}
\label{POGA1}
    S[g]|_{\partial \Sigma}= \int_{\partial \Sigma} \operatorname{Tr}(g^{-1}g')^2/ u' \dd \tau \qquad(X u'\dd \tau= g^{-1} g'\dd \tau)
\end{equation}
on the coset space, where $g\in \mathcal{L}\mathcal{G}$, by identifying $\tilde{b}$ with the Casimir $X^2$ and setting $\xi(\tau)= 1/u'$, for $\dd u =u' \dd \tau$ on $\partial \Sigma \sim S^1$.
The relation between \eqn{schwder} and the boundary action in \eqref{boundcas} can then be understood purely from symmetry considerations.\footnote{By setting $n=1$, from \eqref{coadaction} we get
\be
Ad^*_{\phi^{-1}} (- \frac{t}{24},t) = \left(-\frac{t}{12} \left(\frac{1}{2}{\phi'}^2 + \{\phi(\tau),\tau \}_{\mathcal{S}}\right), t\right)\,.
\ee
With the change of variables $F(\tau)\equiv \tan \left( \frac{1}{2} \phi(\tau)\right)$, the coadjoint action can be written in terms of a single Schwarzian derivative, that is
\be
Ad^*_{\phi^{-1}} (- \frac{t}{24},t) = \left(-\frac{t}{12}  \{F(\tau),\tau \}_{\mathcal{S}}, t\right)\,.
\ee
Therefore, we recognise in \eqref{barnich} the Schwarzian action \eqref{schwder}, while disregarding the contribution of the central extension terms in the pairing. 
} The Schwarzian action is nothing but the action associated to the coadjoint orbits of the Virasoro group with global SL$(2,\mathbb{R})$ symmetry and it can be identified with the particle-on-a-group-manifold action reduced on the coset space. 

\textcolor{black}{A more direct way of deriving the equivalence between the action of a particle on a group manifold \eqref{POGA1} and the natural action on the coadjoint orbits of Diff$(S^1)$ is that of \cite{Valach}, where the group element $g \in$ SL$(2,\mathbb{R})$ is parametrized by means of the Iwasawa decomposition NAK:
\begin{equation}
    g(t) =\begin{pmatrix}
      1 & F \\ 0 &1  
    \end{pmatrix}\begin{pmatrix}
      a^{-1} &0  \\ 0 & a
    \end{pmatrix}\begin{pmatrix}
      \cos{\theta/2} & -\sin{\theta/2}  \\ \sin{\theta/2} & \cos{\theta/2}
    \end{pmatrix}.
\end{equation}
The action functional is then
\begin{equation}
    S[F,a,\theta] =- \int_{S^1} \Big( -\frac{1}{4}\theta'^2 + \Big(\frac{a'}{a}\Big)^2 + \frac{1}{2} a^2 \theta' F'  \Big) d\tau\, .
\end{equation} Conserved charges of the particle on a group action are given by 
\begin{equation}
    J_i = \operatorname{Tr}\big\{T_i g^{-1}g' \big\},
\end{equation}
in particular, one of the conserved charges is the conjugate momentum $\pi_{F}$ of $F$. The explicit computation leads to
\begin{equation}
  J_{-}(g)=  \pi_{F} = \frac{1}{2}a^2 \theta'.
\end{equation}
By means of this constraint, the last term of the action is just $\int F' d\tau$ and therefore vanishes since $F$ is a periodic function. The action in the only residual degree of freedom $\theta$ then coincides with the Schwarzian Action:
\begin{equation}
    S[\theta] = \frac{1}{2}\int_{S^1} \Big( \frac{1}{2} \theta'^2 + \big\{ \theta,\tau \big\} \Big) d\tau.
\end{equation}
See \cite{Valach} for more details.}

In the following, we will seek an extension of the boundary Schwarzian dynamics of JT gravity by extending  the derivation to  a SO$(2,2)$-PSM, which naturally encodes JT gravity  plus extra gauge fields. By first writing the boundary action functional in terms of Casimir functions of the larger group and then using the coadjoint orbit method to compute its reduction, we will see that, under a suitable choice of boundary conditions, we get partial gauge symmetry breaking at the boundary, resulting in a semidirect product structure Diff$(S^1) \ltimes \mathcal{L}G$ for the infinite-dimensional group whose action corresponds to the boundary Casimir action.

\textcolor{black}{The choice of SO$(2,2)$ is motivated by the possibility of providing  a natural gravitational origin to the extra gauge fields emerging in the SO$(2,2)$-PSM, once the gravitational SL$(2,\R)$ sector has been singled out, in terms of a dimensional reduction from a 3d Chern-Simons (CS) theory with boundary~\cite{CSBF}. We briefly recall the dimensional reduction from 3d CS theory to 2d BF/PSM model associated with  JT gravity in the following. Thereby, we move to the proposed SO$(2,2)$ model in Section \ref{Sec3}.} 

\subsection{Dimensional Reduction from 3d CS+WZW}

\textcolor{black}{Let us consider a CS theory in three dimensions, with $\mathfrak{so}(2,2)$-valued connection $\Omega$. This yields    a purely gravitational model describing the $AdS_{3}$ geometry. 
 We shall see that the dimensional reduction of such a theory is equivalent to an $\mathfrak{so}(2,2)$ PSM in two dimensions. Moreover, we will show that the 2d boundary terms, necessary to derive the 1d Schwarzian dynamics, can be recovered by adding to the CS theory a WZW boundary term which dimensionally reduces to the action of a particle on a group manifold.} 

\textcolor{black}{To this,  let $\Sigma_3$ be a manifold with the structure $\Sigma_3= \Sigma \times I $, where $\Sigma$ is a two-dimensional manifold with boundary and $I$ a suitably chosen one dimensional sub-manifold. We write
\begin{equation}
\label{CSaction}
    S_{CS}[\Omega]= \frac{1}{2}\int_{\Sigma_3} \langle\Omega \wedge d\Omega + \tfrac{1}{3}[\Omega, \Omega]\wedge \Omega \rangle.
\end{equation}
It is convenient for the purpose of this section to write $\mathfrak{so}(2,2)$ as the direct sum $\mathfrak{so}(2,2) = \mathfrak{sl}(2,\mathbb{R})_L \oplus \mathfrak{sl}(2,\mathbb{R})_R$. Then $\Omega = \omega^L_i L_i + \omega^R_i R_i $, with $L_i$ and $R_i$ spanning $\mathfrak{sl}(2,\mathbb{R})$. The Killing form is given by
\begin{equation}
    \langle L_i, L_j \rangle = \langle R_i, R_j \rangle =\eta_{ij} , \quad \langle L_i, R_j \rangle =0
\end{equation}
with {$\eta$ = diag(1,-1,1)}. Therefore, the Chern-Simons theory splits into two copies of a single $\mathfrak{sl}(2,\mathbb{R})$-CS theory with connection $\omega = \omega^i\tau_i$. It is then sufficient to prove that each $\mathfrak{sl} (2,\mathbb{R})$-CS sector reduces to a 2d $\mathfrak{sl}(2,\mathbb{R})$-BF theory. The explicit form of each $\mathfrak{sl}(2,\mathbb{R})$ sector of \eqref{CSaction} is given by
\begin{equation}
    S_{CS}[\omega] = \frac{1}{2}\int_{\Sigma_3} d^3x \, \epsilon^{\lambda \mu \nu}\Big( \omega^h_{\lambda} \partial_{\mu} \omega^k_{\nu}+ \frac{1}{3} \omega^{h}_{\lambda} \,f^{k}_{ij}\, \omega^{i}_{\mu} \omega^{j}_{\nu} \Big)  \eta_{hk}
\end{equation}
where $f_{ij}^k$ are structure constants of $\mathfrak{sl}(2,\mathbb{R})$. Let $(\phi, \tau, \rho)$ be local coordinates over $\Sigma_3$, with $\phi \in I$. The dimensional reduction scheme consists in identifying $\omega^i_{\phi} \rightarrow \Phi^i$, $\omega^i_{\tau, \rho} \rightarrow A^i_{\tau, \rho}$ and discarding any derivative with respect to $\phi$, $\partial_{\phi} \rightarrow 0$. This results in the action
\begin{equation}
      S_{CS}[\omega] = \frac{1}{2}\int_{\Sigma_3} d^3x \Big( \omega^h_{\phi}\epsilon^{\mu \nu}\partial_{\mu} \omega^h_{\nu} + \epsilon^{\mu \nu} \omega^h_{\mu} \partial_{\nu} \omega^k_{\phi}   +  \frac{1}{3} \epsilon^{\lambda \mu \nu} \omega^{h}_{\lambda} \,\epsilon^{k}_{ij}\, \omega^{i}_{\mu} \omega^{j}_{\nu} \Big)  \eta_{hk},
\end{equation}
where the $\epsilon^{\mu \nu}$ only refers to the $(\rho, \tau)$ coordinates in $\Sigma$. The first two terms in the action are $\langle\Phi, dA \rangle$ and $\langle A \wedge d\Phi \rangle$, where the external derivative and the wedge product are now performed to be those over $\Sigma$, i.e. in the $(\tau, \rho)$ coordinates. The last term corresponds to $2\langle \Phi, [A,A] \rangle$.
In order to recover a 2d BF theory, we have to recognize $F=dA +[A,A]$. From $d(\Phi A)  = d\Phi \wedge A + \Phi dA$, we deduce that $\langle\Phi, dA \rangle + \langle A \wedge d\Phi \rangle $ = $2\langle\Phi, dA \rangle - \langle d(\Phi, A) \rangle$. One obtains then
\begin{equation}
\label{CSBF2}
     S_{CS}[\omega] = \int_{\Sigma_3} d^3x\, \langle \Phi, F \rangle - \frac{1}{2} \int_{ \partial \Sigma_3} d^2x \,\langle \Phi, A \rangle.
\end{equation} }

\textcolor{black}{In order to make full contact with the 2d dilaton gravity on an AdS$_2$ disk $\Sigma$, we require that, after integrating out the redundant dimension, 
\begin{equation}
    \Sigma_3  \xrightarrow[]{\int_{I} d\phi} \Sigma, \qquad \partial\Sigma_3  \xrightarrow[]{\int_{I} d\phi} \partial \Sigma 
\end{equation} 
Therefore, the dimensional reduction of the CS theory, not only reproduces a BF theory in the  2d bulk $\Sigma$, but also gives a boundary term. In order to also recover the dynamical boundary theory, a boundary term must be added to the 3d action, so that its dimensional reduction, together with the boundary term in \eqref{CSBF2}, leads exactly to the $\mathfrak{so}(2,2)$-PSM with the boundary Casimir.} 
 
\textcolor{black}{The full theory whose dimensional reduction leads to the two-dimensional $\mathfrak{so}(2,2)$-PSM model is a CS theory with the WZW boundary term as in \cite{Carlipso22}. We further refer to \cite{SchwarzOrigin} for details. As explained in \cite{Carlipso22}, the 3d theory has a clear interpretation in terms of BTZ black hole geometry, compatible with the near horizon physics description of JT gravity. The connection between dimensionally reduced AdS$_3$ theories and extremal black holes has been also explored in \cite{Verlinde}.}

\section{The \texorpdfstring{$\mathfrak{so}(2,2)$}{} Poisson Sigma Model}\label{Sec3}

As anticipated in the introduction, our goal is to define a generalized version of JT gravity from a topological gauge theory with a symmetry group containing SL$(2,\mathbb{R})$. The dimensional reduction of pure $AdS_3$-CS theory suggests the $\mathfrak{so}(2,2)$-Poisson sigma model over a two-dimensional manifold $\Sigma = \mathbb{R}\times S^{1}$ as natural candidate in this sense. In this section, we show how isolating the gravitational $\mathfrak{sl}(2,\mathbb{R})$ sector in the case of $\mathfrak{so}(2,2)$ naturally singles out the dynamics of the residual degrees of freedom, which result in additional non-abelian gauge fields that also become dynamical at the boundary.
In particular, we show how the boundary dynamics is encoded in the Casimir functions of the full algebra, analogously to the case of the $\mathfrak{sl}(2,\mathbb{R})$-PSM, and it is naturally reduced on the (Diff$(S^1)\ltimes \mathcal{LG})/\operatorname{SL}(2,\R)$ coset space.

\subsection{The bulk theory}

We start by constructing the bulk theory. Let $\Omega$ be the $\mathfrak{so}(2,2)$-valued connection 1-form over $\Sigma$. The way we decompose the connection and the embedding maps in a given basis of the algebra and its dual is convenient for reasons that will be clear soon. The $\mathfrak{so}(2,2)$ algebra is isomorphic to two copies of $\mathfrak{sl}(2,\mathbb{R})$, so we have a natural chiral basis in $\mathfrak{so}(2,2) \simeq  \mathfrak{sl}_ {L}(2,\mathbb{R}) \oplus \mathfrak{sl}_ {R}(2,\mathbb{R})$ :
\begin{equation}
\label{so222}
    [L_i, L_j] = c^k_{ij} L_k, \qquad  [R_i, R_j] = c^k_{ij} R_k\qquad [L_i,R_j] = 0
\end{equation} 
and with $J_i = L_i + R_i$ we rotate the basis into a 'non-chiral' basis with a $\mathfrak{sl}_{L}(2,\mathbb{R})$ sector invariant under the action of the $\mathfrak{sl}_{J}(2,\mathbb{R})$ sub-algebra 
\begin{equation}
    [J_i, J_j] = c^k_{ij} J_k, \qquad [L_i, L_j] = c^k_{ij} L_k, \qquad [J_i,L_j] = c^k_{ij}L_k.
\end{equation}
 We will refer to the $\mathfrak{sl}(2,\mathbb{R})_{J}$ sub-algebra as the gravitational sector and to the $\mathfrak{sl}(2,\mathbb{R})_{L}$ as the ``non-abelian gauge'' sector. In the non-chiral basis, we then write the  connection $\Omega$ as
\begin{equation}
\label{OJL}
\Omega  = A^i J_i + B^i L_i.
\end{equation}
We denote with $\mathfrak{Z}_i$ the embedding maps $\mathfrak{Z}_i : \Sigma \rightarrow \mathfrak{so}(2,2)^* $ with the Poisson brackets
\begin{equation}
    \{\mathfrak{Z}_i, \mathfrak{Z}_j\} = \Pi _{ij}(\mathfrak{Z}) = f_{ij}^k \mathfrak{Z}_k.
\end{equation}
where $f_{ij}^k$ are the $\mathfrak{so}(2,2)$ structure constants. The corresponding Poisson Sigma model takes the form 
\begin{equation}
\label{PSMOZ}
    S_{P_\sigma} = \int_\Sigma d\Omega_i \wedge \mathfrak{Z}^i + \frac{1}{2}\, \Pi ^{ij}(\mathfrak{Z})\, \Omega_i \wedge \Omega_j. 
\end{equation}
Where we used the invariant bi-linear form 
\begin{equation}
    \langle J_i, J_j   \rangle = \langle L_i, L_j   \rangle = k_{ij}, \qquad \langle J_i, L_j   \rangle = \frac{1}{2}k_{ij},
\end{equation}
with $k_{ij}$ the $\mathfrak{sl}(2,\mathbb{R})$ Killing form. A decomposition similar to that in \eqref{OJL} can be also be performed for the embedding maps. From here on, we denote the dual basis with lowered indices, thus we write
\begin{equation}
    \mathfrak{Z} = \mathfrak{X}_iJ^i + \mathfrak{Y}_iL^i.
\end{equation}
It then follows that
\begin{align} 
\label{nonchiralaction}
\begin{split}
        S_{P_\sigma} = &\int_\Sigma dA_i \wedge \mathfrak{X}^i +\,dB_i \wedge \mathfrak{Y}^i + \frac{1}{2} dA_i \wedge \mathfrak{Y}^i +   \frac{1}{2} dB_i \wedge \mathfrak{X}^i+ \\ &+\frac{1}{2}\, {c_k}^{ij} \, \mathfrak{X}^k \, A_i \wedge A_j\,+ \frac{1}{2}\, {c_k}^{ij} \, \mathfrak{Y}^k \, B_i \wedge B_j \,+\, {c_k}^{ij}  \, \mathfrak{Y}^k \, A_i \wedge B_j \\    &\qquad+\frac{1}{2}\, {c_k}^{ij} \, \mathfrak{X}^k \, A_i \wedge B_j\, + \frac{1}{4}\, {c_k}^{ij} \, \mathfrak{X}^k \, B_i \wedge B_j\, + \frac{1}{4}\, {c_k}^{ij} \, \mathfrak{Y}^k \, A_i \wedge A_j. 
\end{split}
\end{align}
where we can recognize the $\mathfrak{sl}(2,\mathbb{R})_J$ Poisson-Sigma model appearing as the gravitational sub-sector of the theory. The equations of motion for the fields $A_i$ and $B_i$ are
\begin{equation}
\mathfrak{D}_{A} A =0, \qquad \mathfrak{D}_{\Omega} B = c_i^{\,hk}B_h A_kL^i,
\end{equation}
while the equations for the embedding maps read 
\begin{equation}
\delta_{\mathfrak{X}}A=0, \qquad  \delta_ {\mathfrak{Y}}\Omega =-c_i^{\,hk}\mathfrak{X}_hB_kL^i
\end{equation}
This means that the on-shell $A$ field is pure gauge with respect to the $\operatorname{SL}(2,\mathbb{R})_J$ sub-group. On the other hand, the on-shell $\mathfrak{X}$ field is stabilizer for $A$. Notice that the covariant derivative for the $B$ field is computed with respect to the entire connection $\Omega$, while the same is not true for the $A$ field, whose equation of motion is equivalent to the case of the SL$(2,\R)$ BF/PSM  in \eqref{eomf}. This shows that we are allowed to keep interpreting the $A$-sector of the model as equivalent to ordinary JT gravity. Differently, the covariant derivative acting on the $B$ fields has both contributions from $A$ and $B$. Therefore $B$ behaves like a gauge field  coupled to gravity. 

We now introduce a boundary action term and characterize the boundary dynamics in $S^1$. The equation of motion for the gauge connection, that we derive from \eqref{PSMOZ}, is 
\begin{equation}
\label{eomO}
\mathfrak{D}_{\Omega}\Omega = 0
\end{equation}
and, without adding any counterterm, the variation on the boundary is just
\begin{equation}
    \delta S_{P_\sigma}|_{S^1} = - \int_{S^1} \Omega^i \delta \mathfrak{Z}
\end{equation}
As it is the case for the $\mathfrak{sl}(2,\mathbb{R})$-PSM, if we insert a boundary Casimir counter-term  $\mathfrak{Z}^2$ and set the boundary condition 
\begin{equation}
\label{nbc}
\Omega|_{S^1} = \mathfrak{Z}|_{S^1}\dd u
\end{equation}
we get a particle on a group action (cfr \eqref{POGA}) :
\begin{equation}
\label{POGASO22}
    \Omega|_{S^1} = g^{-1}d g \implies S_{P_{\sigma}}|_{S^1}=\int_{S^1}  \frac{1}{2}\operatorname{Tr}\{ (g^{-1} g')^2 \} \frac{1}{u'} \dd \tau,
\end{equation}
where now $g$ is an element in the SO(2,2) gauge group.

We know that the bulk theory has two $\mathfrak{sl}(2,\mathbb{R})_{J,L}$ sectors in interaction and we would like to make that manifest also at the boundary. We expect the boundary action to comprise a Schwarzian derivative term corresponding to the gravitational $\mathfrak{sl}(2,\mathbb{R})_J$-PSM sector  $(\mathfrak{X}^2)$ and a particle-on-a-group term for the non-abelian sector ($\mathfrak{Y}^2$), plus interactions. As we will see, this is indeed the case for a suitable choice of boundary conditions. 

\subsection{Asymptotic Symmetries}
Let's now focus on  the fate of the boundary symmetries and the form of the reduced boundary action. The equation of motion \eqref{eomO} fixes the connection $\Omega$ to be pure gauge
\begin{equation}
    \Omega = g^{-1}dg, \qquad g : \Sigma\rightarrow\operatorname{SO}(2,2)
\end{equation}
This condition, together with \eqref{nbc}, reproduces the same symmetry breaking mechanism of the SL$(2,\R)$-PSM, except for the presence of a global SO$(2,2)$ symmetry in the action \eqref{POGASO22}. The allowed gauge transformation are therefore the SO$(2,2)$ gauge transformations which are constant at the boundary.
Now we show that imposing further boundary conditions can dramatically change the symmetry breaking mechanism and the boundary dynamics can be understood in terms of coadjoint orbits of Diff$(S^1)\ltimes \mathcal{LG}$, for some $\mathcal{G}$ depending on the additional choices made.

Suppose that the embedding fields are chosen at the boundary such that,
\begin{equation}
\label{conXB}
    \mathfrak{X_i}|_{S_1} = -\mathfrak{Y_i}|_{S_1}
\end{equation}
Under these boundary conditions, and \eqref{nbc}, the $\mathfrak{Z}$ field at the boundary is no longer $\mathfrak{so}(2,2)$-valued but $\mathfrak{sl}_R(2,\mathbb{R})$-valued.\footnote{Notice that the choice in \eqref{conXB}, together with the $\Omega|_{S^1}=\mathfrak{Z}|_{S^1}\dd u$ condition, is similar to the gauge fields paramentrisation in the Yang-Mills extension of JT gravity proposed in \cite{Grumill}.} As a consequence, we are reducing the unfixed fields at the boundary.  In fact,
\begin{equation}
    \mathfrak{Z} =  \big( \mathfrak{X_i}+\mathfrak{Y_i} \big)J^i + \mathfrak{Y_i}(L^i - J^i) \xrightarrow[] {\mathfrak{X_i}= -\mathfrak{Y_i}|_{S^1}} \mathfrak{Z}=-\mathfrak{Y}_iR^i
\end{equation}
This additional condition, together with $\Omega$ being pure gauge corresponds to a partial gauge fixing in the $R$-sector. Indeed, also $\Omega$ takes the form 
\begin{equation}
    \Omega|_{S^1} = h_R^{-1}dh_R, \qquad h_R \in \operatorname{SL}_R(2,\mathbb{R})\,.
\end{equation}
% This implies that, by neglecting for the moment the compatibility between gauge transformations and boundary conditions, the SL$_L (2,\mathbb{R})$ is no longer automatically mod-out. 
Moreover, any SL$_L(2,\mathbb{R})$ gauge transformation would not preserve the boundary conditions since it would force the boundary fields to get out of the $R$-sector. Therefore, the SL$_L(2,\mathbb{R})$ gauge group must now be regarded as an actual broken gauge symmetry, while the global SL$_L (2,\mathbb{R})$ symmetry is just trivial since global transformations in this sector are in the Little Group of the boundary connection. The only non-trivial global symmetry is then the SL$_R(2,\mathbb{R})$ symmetry. 
With this in mind, we can conclude that the reduction of the on-shell boundary action involves the coadjoint orbit of Diff$(S^1)\ltimes\mathcal{LG}$, where the given boundary conditions identifies $\mathcal{LG}$ with the SL$_L(2,\mathbb{R})$ loop group. We can then understand the reduced on-shell particle on a group with the action associated to Diff$(S^1)\ltimes\mathcal{L} \text{SL}_L(2,\mathbb{R})$ and SL$_R (2,\mathbb{R})$ global symmetry. %The presence of the Kac-Moody $\mathcal{LG}$ contribution is due to the additional boundary conditions and it strongly depends on their specific form, which in principle could be generalized. 

Since symmetries of the boundary theory are fixed by the additional boundary conditions % beyond the trivial $\Omega|_{S^1} = \mathfrak{Z}|_{S^1}du$. 
\textcolor{black}{it is important  to ask how many nonequivalent choices of boundary conditions exist. Suppose then that the boundary embedding fields are chosen to be valued on a given sub-algebra $\mathfrak{h} \subseteq \mathfrak{so}(2,2)$
\begin{equation}
\label{bcond}
    \mathfrak{Z}|_{S^1} = \mathfrak{H} \in \mathfrak{h}
\end{equation}
Let then $\mathfrak{f}$ be the complement of $\mathfrak{h}$ in $\mathfrak{so}(2,2)$. In order to preserve the boundary condition \eqref{bcond} the following conditions must hold
\begin{align}
    &[\mathfrak{h}, \mathfrak{h}] = 0 \quad or \quad [\mathfrak{h}, \mathfrak{h}] = \mathfrak{h},\\
    &[\mathfrak{f}, \mathfrak{h}] = 0 \quad or \quad [\mathfrak{f}, \mathfrak{h}] = \mathfrak{h}.
\end{align}
By virtue of these conditions, we can list the allowed scenarios in the case of $\mathfrak{so}(2,2)$. The case $[\mathfrak{h},\mathfrak{h}] =0$ is ruled out since $\mathfrak{so}(2,2)$ is semi-simple and it has no invariant abelian sub-algebras.
% and each $\mathfrak{sl}_{L}(2,\mathbb{R})$ has a one-dimensional Cartan subalgebra (CSA), meaning the total CSA of $\mathfrak{so}(2,2)$ is two-dimensional. This means that the biggest abelian sub-algebra in $\mathfrak{so}(2,2)$ is two-dimensional and it can be easily generated by choosing an element $H_1$ in $\mathfrak{sl}(2,\mathbb{R})_{L}$ and an element $H_2$ in $\mathfrak{sl}(2,\mathbb{R})_{R}$. The Lie brackets in each $\mathfrak{sl}(2,\mathbb{R})$ can be written as
% \begin{equation}
%     [H_i, E^{\pm}_{i}] = \pm E^{\pm}_i\qquad [E^{+}_i, E^{-}_{i}] = H_i,
% \end{equation}
% where $E^{\pm}_i$ are the rising and lowering operators,
% hence there is no way the total CSA can be left invariant ($[\mathfrak{f}, \mathfrak{h}] =\mathfrak{h}$) under the action of its complement or commute with it. 
The case $[\mathfrak{h},\mathfrak{h}] = \mathfrak{h}$ can be realized with $\mathfrak{h} = \mathfrak{sl}(2,\mathbb{R})_{L,R,J}$ or trivially with $\mathfrak{h} = \mathfrak{so}(2,2)$ which correspond to the absence of any further boundary condition apart from $\mathfrak{Z}|_{S^1}d\tau = \Omega|_{S^1}$. Any choice associated with $\mathfrak{h} = \mathfrak{sl}(2,\mathbb{R})_{R,L,J}$ or $\mathfrak{h} = \mathfrak{so}(2,2)$ leads to a well defined boundary condition, where $\mathfrak{Z}|_{S^1}$ takes values into a (sub)-algebra which, depending on the specific choice, is either left invariant by its complement or commutes with it. 
}

\subsection{Virasoro-Kac-Moody Coadjoint Orbits}
\label{sec5}
At the end of the last section we showed that additional boundary conditions are allowed at the boundary and imply a partial breaking of the gauge symmetry. This means that if we want to make sense of the boundary action, we have to compute the coadjoint orbit of the Virasoro-Kac-Moody semidirect product as it is the case for SYK-like tensor models \cite{Yoon}. Generally speaking, the computation of coadjoint orbits for semidirect products of infinite dimensional groups is not at all an easy task, but the interesting feature of Diff$(S^1)\ltimes \mathcal{LG}$ is that the Virasoro algebra and the Kac-Moody algebra are related by the Sugawara construction (see Appendix~\ref{appendix}). In particular, the Virasoro algebra acts naturally in a derivative way over the Kac-Moody sector. The algebra $\mathfrak{diff}(S^1)\ltimes \hat{g}$ is in fact given by :
\begin{equation}
\label{virkac}
  \begin{split}
      &[L_m, L_n] = (m-n)L_{m+n} + \frac{c}{12}(m^3-m)\delta_{m,-n}\,;\\
        &[K_{i,m}, K_{j,m}] = c^k_{ij}K_{k, m+n} + m\langle K_i, K_j \rangle \delta_{m,-n}\,;\\
        &[L_m, K_{i,n}]= -nK_{i, m+n}.
  \end{split}
    \end{equation}   
In our case, we have $\hat{g}=\widehat{\mathfrak{sl}(2,\mathbb{R})}$. In order to compute the coadjoint orbit for the semidirect product Diff$(S^1)\ltimes \mathcal{LG}$ we follow \cite{Zuevsky}, where both the Virasoro and the Kac-Moody algebra elements are realized through functions.  Let $(u(\tau),k(\tau),\alpha,\beta)$ be a generic element in the set $\mathfrak{diff}(S^1)\ltimes \widehat{\mathfrak{sl}(2,\mathbb{R})}$, with $u(\tau) \in \mathfrak{diff}(S^1)$, $k(\tau) \in \widehat{\mathfrak{sl}(2,\mathbb{R})}$ and $\alpha, \, \beta$ the respective central elements. Then, a basis independent way to write the commutation relation is simply given by
\begin{equation}
\label{virkac2}
    [(u,k,\alpha,\beta),(v,h,\gamma,\delta)]= ([u,v]_{Vir},[k,h]_{KM} -uh'-vk', \Omega_{Vir}(u''',v), \Omega_{KM}(k,h'))
\end{equation}
where the $\Omega$'s stand for the respective cocycles (defined in details in Appendix \ref{appendix}). 
The group multiplication in Diff$(S^1)\ltimes \mathcal{LG}$ is defines as
\begin{equation}
    (\phi, g)(\psi, h) = (\phi \circ \psi, g.h \circ \phi^{-1})\, ,
\end{equation}
where $\phi, \psi$ are finite diffeomorphisms and $g,h$ are elements of the loop group. In order to compute the coadjoint action of Diff$(S^1)\ltimes \mathcal{LG}$, we must as well introduce the pairing between the algebra and its dual space, which is given by the sum of pairings:
\begin{equation}
    \langle (b,\rho,\alpha,\beta), (v,\lambda, \gamma, \delta)   \rangle = \int b(\tau)v(\tau) d\tau +  \int  \langle \lambda(\tau)\rho(\tau)\rangle_{KM}  d\tau  + \alpha\gamma + \beta \delta.
\end{equation}
Therefore, we have \cite{Zuevsky},
\begin{multline}
\label{coadjoint}
%\begin{split}
Ad^{\,*}_{(\phi, g)}(b,\rho,\alpha,\beta) =( (b\circ \phi) {\phi'}^2 + \alpha\{\phi,\tau\}_{\mathcal{S}} + \langle g^{-1}dg, \rho \rangle \,+ \\
+\frac{1}{2}\beta ||g^{-1}dg||^2 , \phi' (g^{-1} \rho g)\circ \phi + \beta g^{-1}dg, \alpha, \beta).  
%\end{split}
\end{multline}
By pairing the coadjoint orbit with a Lie algebra element we can construct a natural action, as we already saw in \ref{SchActionCoad} for the JT gravity case, and we can finally write the reduced expression for the on-shell action \eqref{POGASO22} :
\begin{equation}
\label{geoaction}
    S|_{S^1} = \int_{S^1} \{ (b\circ \phi) {\phi'}^2 + \alpha\{\phi,t\}_{\mathcal{S}} + \langle g^{-1}dg, \rho \rangle \,+ \\
+\frac{1}{2}\beta ||g^{-1}dg||^2 \} d\tau
\end{equation}
which matches with the results of  \cite{Grumiller}. The  element $b(\tau)$ must be chosen in such a way that the global symmetry be SL$(2,\mathbb{R})$, which is realised through fractional transformations~\cite{Alex}. \textcolor{black}{The geometric action \eqref{geoaction}, without loss of generality, can be thought of as the pairing of $Ad^{\,*}_{(\phi, g)}(b,\rho,\alpha,\beta)$ with a pure Virasoro element and, given the structure of the algebra in \eqref{virkac2}, this can be always rotated in a new element with non vanishing Kac-Moody component. Notice that this would not be true if we paired the coadjoint action with an element of the Kac-Moody subalgebra, because the latter is an invariant sub-algebra. Therefore, this last option could be regarded as a nonequivalent choice that kills the Schwarzian degree of freedom.}

\section{Black Hole Entropy}\label{bhe}
\textcolor{black}{Once the boundary dynamics for the $\mathfrak{so}(2,2)$ PSM has been established with a given boundary condition, it is an interesting check to compute the leading order entropy. Indeed, it is well recognized that in JT gravity the latter can be interpreted as the black hole entropy since the theory admits a gravitational interpretation with the boundary playing the role of a near horizon surface. We shall see that we obtain a consistent result for our model.}

\textcolor{black}{Let $\mathcal{H}[\Omega]$ be the holonomy associated with the  gauge connection $\Omega$
\begin{equation}
    \mathcal{H}[\Omega] = \mathcal{P}\exp{\Big[ -\oint \Omega \Big]},
\end{equation}
the equations of motion fix $\Omega$ to be pure gauge, i.e. $\Omega = g^{-1}dg$. This implies that, if the integration contour is the $S^{1}$ boundary and if we demand for smooth Euclidean solution, then
\begin{equation}
    \mathcal{H} = g(\beta)g^{-1}(0) = \mathbb{1},
\end{equation}
with the inverse temperature $\beta$ being the length of the boundary.}

\textcolor{black}{Suppose that no further boundary conditions are imposed apart from $\mathfrak{Z}|_{S^1}dt = \Omega|_{S^1}$, so that the boundary dynamics is that of a particle on the entire SO$(2,2)$ group. The global SO(2,2) symmetry for the boundary theory implies the presence of the conserved charges
\begin{equation}
     J_i = \langle g^{-1}dg, \tau_i \rangle
\end{equation}
where $\tau_i$ are $\mathfrak{so}(2,2)$ generators. This means that, up to constants, the on-shell connection satisfies $\Omega_i|_{o.s.} = J_i$, therefore 
\begin{equation}
    \mathcal{H}[\Omega] = \exp\Big[ -\beta J \Big] = \mathbb{1}.
\end{equation}
Let now $\Lambda$ be the diagonalized $-\beta J$, i.e. $ -\beta J = P\Lambda P^{-1}$. This implies 
\begin{equation}
    \mathcal{H}[\Omega] = \exp[ P \Lambda P^{-1}] = P \exp [-\beta\Lambda]P^{-1} = \mathbb{1}.
\end{equation}
Requiring $\exp[\Lambda] = \mathbb{1}$ implies that the eigenvalues must satisfy $\lambda_{k} = -2\pi i n_k/\beta$, with integers $n_k$. In the case of an $\mathfrak{so}(2,2)$-valued boundary connection, the 4 eigenvalues come in two pairs of eigenvalues with opposite sign: 
\begin{equation}
    \operatorname{Tr}(J^2) = -\frac{8\pi(n^2 +m^2)}{\beta^2}.
\end{equation}
In order to evaluate the entropy we need to compute the leading order free energy $F$ which can be identified with the temperature times the on-shell boundary Euclidean action $I$
\begin{equation}
    F = \beta^{-1} I|_{S^1, \, on-shell}.
\end{equation}
The latter can be computed easily. Let $\mathcal{C}$ be the on-shell boundary Casimir function, then $\mathcal{C} = \langle J,J \rangle$ and
\begin{equation}
    I|_{S^1, \, on-shell} = \beta \mathcal{C} = \beta \operatorname{Tr}(J^2) = -\frac{8\pi (n^2+m^2)}{\beta}.
\end{equation}
The free energy then comes with the correct sign thus recovering a positive entropy which is linear in the temperature $T=\beta^{-1}$, that is
\begin{equation}
    S = -\frac{dF}{dT} = 16 \pi k (n^2 + m^2)T
\end{equation}
This result is consistent with the linear scaling of the entropy in SYK models \cite{Maldacena_2016, Grumiller} and it shows that the extra SL$(2,\R)$ degrees of freedom contribute with a supplementary linear term $\propto \beta^{-1} m^2$.   
}

\section{Conclusions}
\label{Sec4}
%\subsection{Summary}
\textcolor{black}{We constructed a JT gravity bulk-plus-boundary generalization in terms of a $\mathfrak{so}(2,2)$-Poisson sigma model with a boundary Casimir action. It is a fact that 3d Chern-Simons theories with WZW term at the boundary, once dimensionally reduced, give  2d BF theories with the particle on a group action at the 1d boundary~\cite{SchwarzOrigin}. The dimensional reduction of SO(2,2) Chern-Simons-WZW theory, which is a 3d theory describing an AdS$_3$ geometry, leads to the proposed $\mathfrak{so}(2,2)$-PSM together with the dynamical particle on a group action.\footnote{In three spacetime dimensions, $\mathfrak{so}(2,2)$ is in fact a kinematic algebra in the sense of  \cite{Kine}, and a 3d Chern-Simons theory over this algebra provides a purely gravitational theory of AdS$_3$ geometry.}
The model provides a gravitational dual for SYK-like tensor models with {internal} symmetries \cite{Yoon, NarayanYoon}, whose low energy dynamics is characterized by a $\mathfrak{diff}(S^1)\ltimes \Hat{\mathfrak{g}}$ symmetry. In our case, the $\mathfrak{diff}(S^1)\ltimes \Hat{\mathfrak{g}}$ symmetry, with $\mathfrak{g} = \mathfrak{sl}(2,\mathbb{R})$, arises after a specific choice of boundary conditions. The additional Kac-Moody sector encodes the dynamics of extra edge modes, whose number is equal to $\operatorname{dim}(\mathfrak{g})$. In the near horizon interpretation, the standard computation of the near-extremal black hole entropy reproduces the JT result with an additional entropy contribution due to the presence of extra gauge fields. The additional contribution is equivalent to the bare JT case because of the particular structure of $\mathfrak{so}(2,2)$.
The specific case of $\mathfrak{so}(2,2)$, seen from the CS perspective, allows to interpret the Kac-Moody modes, living at the $S^1$ boundary and expected to arise in the low-energy regime of SYK tensor models, as Kaluza-Klein modes associated with the dimensional reduction.
The proposed $\mathfrak{so}(2,2)$-PSM provides a class of possible JT-Yang-Mills generalizations of the JT/SYK correspondence with a purely gravitational interpretation from a 3d perspective. This interpretation selects a class of SYK-like duals which naturally relates to 3d gravity.} 

\textcolor{black}
{The choice of working with a PSM, which in the present paper  is linear, therefore equivalent to a BF theory up to boundary therms, lends itself to generalizations which could have interesting gravitational implications. The first generalization is related with  the possibility of considering non-linear PSM; this has already been noticed in the literature  \cite{Grumiller:2021cwg}, it being related to nonlinear generalizations of JT gravity \cite{Ikeda:1993fh, IKEDA1994435}.}

\textcolor{black}{The second generalization, entirely novel up to our knowledge, would be to  consider  Jacobi sigma models \cite{Bascone:2020drt, Chatzistavrakidis:2020gpv, Bascone:2021njt, DiCosmo:2024jum}, which are models with a deformed Poisson bracket on the target space, known as Jacobi bracket. The latter is constructed with a quasi-Poisson structure, violating   Jacobi identity in a controlled manner.  The explicit relation for the quasi-Poisson tensor $\Pi$ reads $[\Pi, \Pi]= 2 E\wedge\Pi$, with  $E$ a vector field on the target space,  the so-called Reeb vector field, s.t. $L_E \Pi= 0$. This structure may be obtained by a homogeneous Poisson bracket in one dimension higher (technically a fiber bundle over the target space, with fiber $R-\{0\}$ \cite{DiCosmo:2024jum}). Interestingly, the extra dimension, or, better to say, the generator of  $R-\{0\}$, could be related  with a  dilaton field   which could play a role in generalized JT gravity models. We plan to analyse in detail this proposal in future investigations. }

\appendix

\section{Appendix}
\label{appendix}
\subsection{The Sugawara Construction}\label{suga}
Here we summarize the basics of the Sugawara construction and give a more detailed computation of the Virasoro coadjoint orbits. 

Given a semi-simple finite-dimensional Lie algebra $\mathfrak{g}$, and the  algebra of smooth functions on the circle $C^{\infty}(S^{1})$, we denote the corresponding loop algebra as 
\begin{equation}
    L{\mathfrak{g}} = \mathfrak{g}\otimes C^{\infty}(S^{1}).
\end{equation}
Given $\bf{x} \in \mathfrak{g}$, and  $\{\e^{\ii nt}\}_{n\in \Z}$ a basis  for  $C^{\infty}(S^{1})$,  a basis for $L\mathfrak{g}$ is represented by  $\{X_n\}$,  so defined  
 \begin{equation}
 X_{n} = {\bf x}\,e^{\ii nt}
 \end{equation}
with Lie brackets 
\begin{equation}
    [X_{m}, Y_{n}] = [{\bf x},{\bf y}]\,\e^{\ii (m+n)t} =: X_{m+n}  \label{loop}
\end{equation}
%and $[{\bf x},{\bf y}]_{m+n}:= [{\bf x},{\bf y}]\,\e^{\ii (m+n)t}$.
%where the commutator stands for the Lie Parenthesis in $\mathfrak{g}$ and the product over $C^{\infty}(S^{1})$ is just the point-wise product of functions. Now we have to perform a central extension of the Loop Algebra and we will improperly refer to this new structure as a \textbf{Kac-%Moody Algebra}. 
Central extensions of $L\mathfrak{g}$  may be obtained by considering projective representations \cite{Wassermann} $\pi: X_m\mapsto \pi(X_m)=:X(m)$ s.t. $[\pi(X), \pi(Y)]= \pi([X,Y])+ \alpha({X},{Y}){\bf e}$, where $\alpha({X},{Y})\in \R$ is a a two-cocycle on $L\mathfrak{g}$, namely  a bilinear map from $L\mathfrak{g}\times L\mathfrak{g}$ to the complex numbers, which is antisymmetric   and verifying the cocycle property 
\be\label{Omegacycl}
   \del\alpha(X,Y,Z):= \alpha(X,[Y,Z])+ \alpha(Z,[X,Y]) + \alpha(Y,[Z,X]) = 0
\ee
while ${\bf e}$ is the generator of a one-dimensional vector space extending the finite Lie algebra $\mathfrak{g}$. The representation can be chosen in such a way to fix the value of the cocycle, so to obtain
%${K} \in  
%\{\mathfrak{a}\}\otimes C^{\infty}(S^{1})$, with $\mathfrak{a}$ an Abelian algebra commuting with  $\mathfrak{g}$, which commutes with all others generators of the loop algebra
%$\begin{equation}
%    [{K}, X_{n}] = 0 \qquad \forall n \in \mathbb{Z}.
%\end{equation}
%We call this new generator the {\it center}. Of course, now we have to %fix a new algebra structure. 
\begin{equation}
\label{kacalg}
    [X(m), Y(n)] = \pi(X_{n+m}) + m \ell\,\langle {\bf x}, {\bf y} \rangle \delta_{n+m,\,0}
\end{equation}
with $\langle \, , \, \rangle$ the (non-degenerate, invariant) inner product in the semisimple algebra $\mathfrak{g}$; $\ell$ is a non-negative integer number determined by the dimension of the representation.
Hence, $\mathcal{L}\mathfrak{g}= L\mathfrak{g}\oplus \R{\bf e}$
is a central extension of the loop algebra $L\mathfrak{g}$,  what is called an affine Lie algebra. Note that $\R{\bf e}$ is a central subalgebra of $\mathcal{L}\mathfrak{g}$. 

By denoting  with ${\x}_i, i=1,\dots,{\rm dim} \,\mathfrak{g}$, an orthonormal basis in $\mathfrak{g}$ and $X_i(p)= \pi(\x_i \, \e^{\ii p t})$ an orthonormal basis in $\mathcal{L}\mathfrak{g}$, the Lie brackets of the affine algebra  \eqn{kacalg} acquire the form
\be\label{normalform}
[X_i(m), X_j(n)] = f_{ij}^k X_k(m+n) + m \ell\,\delta_{ij} \delta_{n+m,\,0}.
\ee
 It is also useful to introduce the current basis\footnote{Notice that $\tau$ can also be chosen on the real line, with appropriate boundary conditions.}
 \be\label{currba}
 \mathcal{X}_i(\tau)= \sum_n \e^{\ii n \tau} X_i (n), \quad \tau\in S^1
 \ee
It can be checked that the current basis satisfy
\be\label{curra}
[\mathcal{X}_i(\tau), \mathcal{X}_j(\tau')]=  f_{ij}^k \mathcal{X}_k(\tau) \delta(\tau-\tau') + \delta_{ij} \delta' (\tau-\tau') \ell
\ee
namely, $\mathcal{X}_i$ are maps from the circle $S^1$ to the algebra $\mathfrak{g}$, which are properly called the generators of the loop algebra associated with $\mathfrak{g}$ (see \cite{Goddard} for details). The latter shall be  indicated with $\mathcal{L}\mathfrak{g}$ and is often referred to as the current or Kac-Moody algebra of $\mathfrak{g}$.

From $\mathcal{L}\mathfrak{g}$  a Virasoro algebra is constructed using the Sugawara construction \cite{Sugawara, Goddard}. 
%To this, by denoting  with ${\x}_i, i=1,\dots,{\rm dim} \,\mathfrak{g}$, an orthonormal basis in $\mathfrak{g}$ and $X_i(p)= \pi(\x_i \e^{\ii p t})$ an orthonormal basis in $\mathcal{L}\mathfrak{g}$, one defines 
The Virasoro generators are defined according to 
\begin{equation}\label{Vira}
    T_m = -\frac{1}{2(\ell+h)}   \sum_{i} \sum_{p+q = m} :X_{i}(p)X_{i}(q): 
\end{equation}
where $h$  is the dual Coxeter number of $\mathfrak{g}$\footnote{$h=1/{\kappa}^*(\theta_C,\theta_C)$ with $\kappa^*$ the scalar product in the dual of the Cartan subalgebra of $\mathfrak{g}$ derived from the Killing metric.}. They satisfy 
\be\label{TT}
 [T_m, T_n] = (m-n)T_{m+n} + \frac{C}{12}(m^3 - m)\mathbf{e}\,\delta_{n+m, \, 0}
 \ee
 with   $C={\rm{dim}}\,\mathfrak{g}\cdot \ell/(\ell+ h)$ (see \cite{Wassermann} for a  proof of \eqn{TT}). We shall indicate the Virasoro algebra as $\mathcal{V}= {W}\oplus \R\mathbf{e}$, with ${W}$ the Witt algebra\footnote{More precisely, the Witt algebra is the algebra of smooth vector fields on the circle, to which the algebra \eqn{TT} with $C=0$   is isomorphic. We shall use the same name to simplify the notation}.
 
 %\textcolor{magenta}{Questa base dovrebbe essere ortonormale rispetto a un qualche prodotto scalare indotto dalla loop algebra. Sarebbe utile verificarlo, perche' e' quello che si usa per dimostrare che il cociclo A.17 si riduce a quello in A.10 quando valutato sui generatori}
 
The Sugawara tensor is then defined in terms of the currents according to \be\label{Sugaten}
{T}(\tau) = \frac{1}{2(\ell+h)} \sum_i : \mathcal{X}_i(\tau) \mathcal{X}_i(\tau).
\ee
It is a standard result of the Sugawara construction in two-dimensional quantum field theory  to verify that it satisfies the Lie brackets
\be\label{sugabra}
[T(\tau), T(\tau')]= \big({T}(\tau)+{T}(\tau')\bigr)\delta'(\tau'  - \tau) - \frac{C}{12}\big(\delta'''(\tau -\tau')+\delta'(\tau -\tau') \big). 
\ee
Hence, it can be verified that the generators of the Virasoro algebra \eqn{Vira} are the Fourier  modes of the Sugawara tensor \eqn{Sugaten}.Finally, notice that the quadratic operator
\be
\Delta=\sum_i :X_i(0) X_i(0):
\ee
is a Casimir for the affine algebra $\mathcal{L}\mathfrak{g}$ \eqn{kacalg}, or equivalently, the quadratic operator
\be
\sum_i :\mathcal{X}_i(\tau) \mathcal{X}_i(\tau):
\ee
is a Casimir for the same algebra, w.r.t. the bracket \eqn{curra}.

\subsection{Coadjoint Orbits}\label{A2}
%So far we introduced the concept of Central Extension in a basis-dependent way. A more general overview can be the following. A central exension of a Lie algebra $\mathfrak{g}$ is basically $\hat{\mathfrak{g}}\oplus \mathbb{R}$. Therefore, an element in $\hat{\mathfrak{g}}$ is a couple $(v,x)$ where $v \in \mathfrak{g}$ and $x$ is the central component. We need to introduce a new Lie Parenthesis over this new objects. This corresponds to enlarge the structure constants of $\mathfrak{g}$ in order to accommodate for the new central element.It is easy to realize that this additional structure constant have two indices in a $basis-dependent fashion (see \ref{kacalg}). Since any generator of the centrally-extended algebra commutes with the center by definition, we deduce that this new 2-indices structure depends only on the Lie Algebra compontents of the $(v,x)$ couples.% 
The Kac-Moody and Virasoro algebras introduced in the previous section can be recast in a basis independent fashion, which allows for a direct computation of the coadjoint orbits of their groups.   
Let us consider the affine algebra first. Given $(v,a), (u,b)$ two elements in $\mathcal{L}\mathfrak{g}= L\mathfrak{g}\oplus \R \mathbf{e}$,  with $v, u \in L\mathfrak{g}$ and $a,b  \in \R \mathbf{e}$,   the general form of the commutation relations is:
\begin{equation}\label{vu}
    [(u,a),(v,b)]_{\mathcal{L}\mathfrak{g}}= \left([u,v]_{L\mathfrak{g}}, c(u,v)\right)
\end{equation}
where $c$ is a two cocycle on the Lie algebra $L\mathfrak{g}$ (cfr. \eqn{Omegacycl}).
We recall that the inequivalent non-trivial central extensions of $L\mathfrak{g}$ are classified by equivalence classes of 2-cocycles (see for example \cite{Schottenloher}). 
It is easily verified that \eqn{vu} reduces to \eqn{normalform},  when computed on the generators $X_i(m)$. The cocycle in \eqn{vu} can be written in integral form according to 
\begin{equation}
    c(u,v) =  \frac{1}{2\pi}
    \oint \dd \tau\, \langle u'(\tau),\,v(\tau)\rangle \,
\end{equation}
with $\langle\cdot\,,\cdot\,\rangle$ the inner product in $L\mathfrak{g}$.
It is immediate to check  that it coincides with the central term of \eqn{normalform} when computed on the currents \eqn{currba}. Analogously, an integral expression can be obtained for the Virasoro cocycle. To this (see for example \cite{wittencoad, Valach}) one has to represent the Virasoro algebra $\mathcal{V}= {W}\oplus \R\mathbf{e}$ in terms of vector fields on the circle $\xi= \xi(\tau) \del_\tau\in W$ plus a central term. Then, for $(\xi,\alpha),(\eta,\beta)\in \mathcal{V} $ one has 
\be \label{a16}
 [(\xi,\alpha),(\eta,\beta)]_{\mathcal{V}}= \left([\xi,\eta]_{W}, c_0(\xi,\eta)\right)
 \ee
with $\alpha,\beta\in \R$ and 
\be
[\xi,\eta]_{W}= \bigl(\xi(\tau)\del_\tau \eta(\tau)-\eta(\tau)\del_\tau \xi(\tau)\bigr)\del_\tau
\ee
while the two-cocycle is given by:
\begin{equation}\label{Omegacoh}
    c_0(\xi,\eta) = \frac{\ii t}{48\pi}\oint\, \dd \tau \, \bigl(\eta'''(\tau)\xi(\tau) - \eta(\tau)\xi'''(\tau) \bigr), \quad\quad t\in\R
\end{equation}
known as the Gel'fand-Fuchs cocycle. In order to check that it  reduces to the central term in \eqn{TT} one has to use the realization of the Witt algebra in terms of  the generators $\xi_n= f_n(\tau) \del_\tau$ with   $f_n = \exp(\ii n \tau)$. Then one has  
\be
c_0(\xi_n, \xi_m)= \frac{\ii t}{48\pi}\oint \dd \tau \bigl(\xi_n^{'''}(\tau) \xi_m (\tau)- \xi_n(\tau) \xi_m^{'''} (\tau)\bigr)= \frac{t}{12}m^3\delta_{n+m,0}
\ee 
which is cohomologous to the two-cocycle in \eqn{TT}\cite{Grabowski:1996sb}. Indeed, in order to obtain the cocycle in \eqn{TT} one has to modify \eqn{Omegacoh} by a coboundary term as follows
\begin{equation}\label{Omegacohom}
    c (\xi,\eta) = \frac{\ii t}{48\pi}\oint\, \dd \tau \, \bigl((\eta'''(\tau)-\eta'(\tau))\xi(\tau) - \eta(\tau)(\xi'''(\tau) -\xi'(\tau))\bigr)
\end{equation}
it being
\be\label{cobound}
\oint\, \dd \tau \, \bigl((\eta'(\tau))\xi(\tau) - \eta(\tau)\xi'(\tau))\bigr)= \del b(\xi,\eta)=b([\xi,\eta])
\ee
with $b$ a one-cochain in the Gel'fand-Fuchs cohomology. Once the central extensions are defined, in order to compute coadjoint orbits of the corresponding groups, it is necessary to introduce the notion of coadjoint action on the dual algebras. This may obtained in a standard way, by first computing the adjoint action on the algebra and then computing its dual through the natural pairing between algebra and dual algebra. We give here a short review while referring  to \cite{Witten, wittencoad} for more details. The adjoint action of the loop group on its algebra is given by
\begin{equation}
\label{AdKM}
    \widetilde{Ad_g}(v(\tau),a) = \left(\, Ad_g v(\tau), a + c(g^{-1}g', v(\tau)\right)
\end{equation}
with $g: S^{1}\rightarrow G$ an element of the affine group, so that $g^{-1}g'\in L\mathfrak{g}$. The tilde indicates the action of the affine group on its centrally the extended algebra, while $Ad_g$ is the standard action. Notice that the adjoint action only affects $v(\tau)$, $a$ being central. Hoewer, the adjoint action on $v(\tau)$  contributes a non central  and a central  term which adds to $a$. 
 On computing \eqn{AdKM}  on a Lie algebra generator $(X(m), a) $, one can check that  
 its  infinitesimal version  reproduces  the adjoint action of the algebra, Eq. \eqref{vu}, it being $g^{-1}g'= \sum c_n X(n)$. 
 Furthermore, one can verify  that \eqn{AdKM} satisfies the ordinary composition property
\be \widetilde{Ad}_{gh} = \widetilde{Ad}_{g} \circ \widetilde{Ad}_{h}. \ee
The adjoint action of the Virasoro group on its algebra is more subtle. Analogously to the previous case, the action on the central term is trivial, so to have 
\begin{equation}
\label{AdVir}
    \widetilde{Ad_g}(v,a) = \left( [Ad_g v]_{nc}, a + [{Ad_g} v]_c\right)
\end{equation}
As for the contribution of ${Ad_g} v$ to the central term, it is possible to 
show  that it can be formulated in terms of Schwarzian derivative according to 
\be\label{advc}
[Ad_g v]_c=\frac{1}{12}\int \frac{\dd \tau}{2 \pi} \frac{v(\phi(\tau))}{\phi'(\tau)} \{\phi,\tau \}_\mathcal{S}
\ee
with 
\be \label{Schwarzian}
   \{\phi(\tau), \tau \}_\mathcal{S} = \frac{\phi'''}{\phi'} - \frac{3}{2}\left( \frac{\phi''}{\phi'}\right)^2 
\ee
To this, we shall follow Kirillov \cite{Kirillov} and shortly review  his  theory of coadjoint orbits. 
%The relationship between Schwarzian derivative and cocycle still awaits clarification. 
Let us first  fix the notation. Given $G=\text{Diff}(S^1)$ and its Lie algebra  of smooth vector fields on the circle, $W$,  the dual algebra $W^*$ consists of continuous linear functionals on $W$. Therefore, it can be identified with $\Omega(S^1)\otimes \mathcal{D}'(S^1)$ with 
$\mathcal{D}'(S^1)$ the space of continuous linear functionals on $\mathcal{F}(S^1)$. Namely, the elements of $W^*$, called moments by Kirillov, 
are of the form
\be\label{moment}
p= f\otimes F\in \Omega^1(S^1)\otimes \mathcal{D}'(S^1),\quad\quad
(f\otimes F)(\xi)=\langle F,f(\xi)\rangle
\ee
with $\xi\in W$. Therefore, $\mathcal{D}'(S^1)$ can be further identified with $\Omega^1(S^1)$ itself, since it provides the volume form to be used in order to integrate $f(\xi)\in \mathcal{F}(S^1)$ in \eqn{moment}. In coordinat form we have
\be 
p= f\otimes \omega= f(x) dx\otimes \omega(x) dx, \quad\quad x\in S^1.
\ee
Then, given $\phi\in \text{Diff}(S^1)$, $\phi(x)=y$, the coadjoint action of the group on $W^*$ is so defined
\be
Ad^*_\phi p= f(\phi^{-1}(y) \frac{\del \phi^{-1}}{\del y} dy\otimes \omega(\phi^{-1}(y)) \frac{\del \phi^{-1}}{\del y} dy
\ee
while the coadjoint action of the Lie algebra $W$ on $W^*$ is given by the Lie derivative
\be
ad_\xi^* p= L_\xi f\otimes \omega + f \otimes L_\xi \omega.
\ee
The Virasoro group is the centrally extended group of diffeomeorphisms of the circle, $\widehat{\text{Diff}(S^1)}$, with product rule
\be
(\phi,t)\circ (\psi,s)= (\phi\circ\psi, t+s+ B(\phi, \psi))
\ee
with $B$ a 2-cocycle on $\text{Diff}(S^1)$, whose explicit form can be found for example in \cite{Kirillov}. The Lie algebra of $\widehat{\text{Diff}(S^1)}$ is the Virasoro algebra $\mathcal{V}= W+ \R\mathbf{e}$ with Lie bracket \eqn{a16}.
The Gel'fand Fuchs cocycle \eqn{Omegacoh} forms a basis in $H^2(W, \R)=Z^2(W,\R)/B^2(W,\R)$, the second cohomology group of the Witt algebra. A representative of the equivalence class, $c_0\in Z^2(W,\R)$ can be written in the form
\be
c_0(\xi,\eta)= \oint \xi' d\eta'
\ee
and a generic element in the class has the form
\be
c(\xi,\eta)= t c_0(\xi,\eta)+p ([\xi,\eta])
\ee 
where $t\in \R$ and $p\in W^*$, namely of the form \eqn{moment} ($c=c_0+ d p$ with $dp$ a coboundary). Prior to compute the coadjoint action of the Virasoro group on its dual algebra, we state without proof the following theorem \cite{Kirillov}
\begin{theorem}
Given $W$, $\mathcal{V}$ and their duals, it is possible to exhibit bijective maps 
\beqa\label{diagram}
\alpha: W^*&\rightarrow & B^2(W)\nonumber\\
\beta: \mathcal{V}^*&\rightarrow & Z^2(W)\nonumber\\
\gamma: \R &\rightarrow & H^2(W)\nonumber
\eeqa
which are $G$-module morphisms,\footnote{A $G$-module is an Abelian group, $A$, on which $G$ acts respecting   the Abelian  structure. A $G$-module morphism $\delta$ is a group homomorphism: $\delta(g(a+b))= \delta(g(a)+g(b))= g(\delta(a+b)), g\in G, a,b\in A$. All of the groups in \eqn{diagram} are obviously $G$-modules for the group of diffeomorphisms.} with $G= \operatorname{Diff}\, S^1$.
\end{theorem}
The maps $\alpha, \beta, \gamma$ are explicitly chosen to be
\beqa
\alpha(f)(\xi,\eta)&=&\langle f,[\xi,\eta]\rangle\\
\beta(f,t)(\xi,\eta)&=&\langle f,[\xi,\eta]\rangle-t c(\xi,\eta)\\
\gamma(t)&=& {\rm class}\, t\,c .
\eeqa
What is interesting for us is the statement for the map $\beta$, which implies in particular that it  is a $\mathcal{V}$-module morhism, with $\mathcal{V}$ the Virasoro algebra. This amounts to the result
\be
\beta\bigl(\widetilde{ad^*}_{(\xi,\tau)}(f,t)\bigr)= \widetilde{ad^*}_{(\xi,\tau)}\bigl(\beta (f,t)\bigr)
\ee
Let us prove this equality explicitly. We have 
\beqa
\langle(\widetilde{ad^*}_{(\xi,\tau)}(f,t)),(\eta,\sigma)\rangle&=& \langle (f,t), \widetilde{ad}_{(\xi,\tau)}(\eta,\sigma)\rangle= \langle (f,t), ([\eta,\xi], c(\eta,\xi)) \rangle \nonumber\\
&= &\langle f, ([\eta,\xi]\rangle - t c(\eta,\xi)
\eeqa 
Recalling that $c$ is a a two-cocycle on $W$, the Lie derivative and the interior product  act as follows
\begin{equation}
\label{coad2}
    \langle i_{\xi}c, \eta \rangle = c(\eta,\xi), \qquad (L_{\xi}c)(\eta, \chi) = ((i_\xi d + d i_\xi)c)(\eta,\chi)= c([\eta,\chi], \xi)
\end{equation}
where $ df(\xi,\eta)= \langle f, [\xi,\eta]\rangle$  and $dc= 0$ have been used. Hence  we obtain 
%From \eqref{coad} and the latter definitions it immediately follows that
\begin{equation}
\label{coad3}
    \widetilde{{ad}^*}_{\xi}(f,t) = (ad^*_{\xi}f  -ti_{\xi}c, 0) 
\end{equation}
%where we have identified the cocycle $c$ with the corresponding element in the dual algebra, as in \eqn{coad2}. Moreover, for $f\in \mathfrak{g}^*$, we have
Moreover, for $f\in \mathfrak{g}^*$ we have  
\be \label{coadd}
ad^*_\xi f= L_\xi f
\ee
then 
\be\label{dcoad}
d({ad}^*_{\xi}f - ti_{\xi}c) = L_\xi (df -t c)
\ee
which is in turn equal to $L_\xi (\beta (f,t))$, that is  what we wanted to show. From \eqn{dcoad} it is possible to obtain the full coadjoint action of the Virasoro group \cite{Kirillov}. By denoting with $L(\phi)$ the one parameter group generated by $L_\xi$ and observing that $L(\phi) c$ generates a 2-cocycle, say  $\tilde c$, which is in the same cohomology class as c,
one has
\be \label{coboun}
L(\phi) c = c+ dh(\phi) 
\ee
with $dh(\phi)$ a coboundary. Thus, on integrating \eqn{dcoad}, we get 
\be\label{coadfin}
\widetilde{{Ad}^*}_\phi (f,t)= (L(\phi) f+t  h(\phi), t).
\ee
We can now explicitly prove that $h=S(\phi)$ when $c$ is the Gel'fand-Fuchs cocycle.
\beqa
 L(\phi) c_0(\xi,\eta)-c_0(\xi,\eta)&= &  c_0(Ad_{\phi}^{-1}\xi, Ad_{\phi}^{-1}\eta) -  c_0(\xi,\eta) \nonumber\\
 &=&\int \Big(\frac{\xi(\phi(\tau))}{{\phi'}(\tau)} \Big)' d\Big( \frac{\eta(\phi(\tau))}{{\phi'}(\tau)}\Big)' -\int \xi'(\tau)\eta''(\tau)\dd\tau.
\eeqa
After the change of variable $\tau \rightarrow \phi^{-1}(\tau)$ in the first integral we get
\begin{equation}
     L(\phi) c_0(\xi,\eta)-c_0(\xi,\eta) = \int  [\xi \eta'-\xi'\eta]\Big( \frac{{\{\phi(\tau),\tau \}_{\mathcal{S}}}}{{\phi'}^2}\Big) \circ \phi^{-1} \,\dd\tau
\end{equation}
where $\{\phi(\tau),\tau \}_{\mathcal{S}}$ is  the Schwarzian derivative \eqn{Schwarzian}. 
On comparing this result with \eqn{coboun} we have
\be
dh(\phi)(\xi,\eta)= \langle h(\phi),([\xi,\eta]\rangle= \int  [\xi \eta'-\xi'\eta]\Big( \frac{{\{\phi(\tau),\tau \}_{\mathcal{S}}}}{{\phi'}^2}\Big) \circ \phi^{-1} \,d\tau
\end{equation}
hence the functional $h$ is explicitly given by,
\be\label{hphi}
h(\phi)= d\tau\otimes \{\phi(\tau),\tau \}_{\mathcal{S}}\circ\phi^{-1} d\tau \in \mathcal{W}^*.
\ee
In order to better understand its meaning, we recall that $W^*\equiv \Omega (S^1)\otimes \Omega( S^1)$ (cfr. \eqn{moment}), where the first differential realises the pairing with vector fields, whereas the second furnishes the integration measure on the circle. This explains the interpretation of the Schwarzian 
\be
S(\phi):= d\tau\otimes \{\phi(\tau),\tau \}_{\mathcal{S}} d\tau
\ee
as a pseudometric on the circle \cite{Kirillov}.
Finally, replacing the result in \eqn{coadfin},  the coadjoint action of the Virasoro group on its dual algebra reads 
\begin{equation}\label{caodVir}
    \widetilde{Ad^*}_\phi (f,t) =(Ad^*_\phi f+ t S(\phi)\circ\phi^{-1},t)= ((f+ t S(\phi))\circ\phi^{-1},t)
    %{\phi'}^2 b(\phi(\tau)) - \frac{t}{12}\{\phi(\tau),\tau \}_{\mathcal{S}}.
\end{equation}
The homogeneous spaces defined by the coadjoint action  are given by the quotient $\operatorname{Diff(S^1)}/\operatorname{Stab}(f)$.
The stabilizer can be computed in some cases and for some constant values of $f(\tau)$. In particular, we have that for $f(\tau) =- \frac{cn^2}{24}$
\begin{equation}
\operatorname{Stab}(b) = \operatorname{Sl}^{(n)}(2, \mathbb{R}).
\end{equation}
where $\operatorname{Sl}^{(n)}(2, \mathbb{R})$ denotes the $n$-fold cover of $\operatorname{Sl}(2, \mathbb{R})$ (see e.g. \cite{Valach}).

\section*{Acknowledgments}
The authors acknowledge support from the INFN Iniziativa Specifica GeoSymQFT and from  the European COST Action CaLISTA CA21109.  P.V. acknowledges support from the  Programme STAR Plus, financed  by UniNA and Compagnia di San Paolo, and from  the PNRR MUR Project No. CN 00000013-ICSC.

\bibliographystyle{utphys}
\bibliography{refs}

\end{document}